# An analogy between four parametrizations of the dark energy equation of state onto Physical DE Models

Ehsan Sadri[*]

*Physics Department, Azad University Central Tehran Branch, Tehran, Iran*

## Abstract

In order to scrutinize the nature of Dark Energy, many Equations of State have been proposed. In this context we examine four popular models (the Chevallier-Polarski-Linder (CPL) model, Jassal model, Barboza & Alcaniz model and Wetterich model) in the field of quintessence and barotropic dark energy models, to compare with each other. We develop a theoretical method of constructing the quintessence potential directly from the four equations of state which describes the properties of the dark energy.

[*] Email: ehsan@sadri.id.ir

# Introduction

From 1929, all cosmologists were studying on the rate of expansion or constriction of the universe. They were of the opinion that the universe with its powerful gravity must be constricting. But Alexander Friedmann and Edwin Hubble showed that the universe is expanding with accelerating rate. Originated from Friedmann's Equations and Hubble constant, Observations data of the accelerated expansion of the universe, indicate that almost 68.3% of the energy density in the universe is in the form of a component which has negative pressure (exactly against the gravity), called dark energy, with the remaining 31.7% in the form of nonrelativistic matter (including both baryonic matter and dark matter). The dark energy can be parametrize by its equation of state parameter which in its general form is:

$$\omega = \frac{P}{\rho}, \qquad (1)$$

The simplest form for the dark energy is a cosmological constant $\Lambda$, which has pressure $P_\Lambda = -\rho_\Lambda$. Specifically, a constant model should describe how the present amount of the dark energy is so small compared with the fundamental scale (fine-tuning problem) and why it is comparable with the critical density today (coincidence problem). The cosmological constant is affected by both these problems. One possible approach to constructing a feasible model for dark energy is to associate it with a slowly evolving and spatially homogeneous scalar field $\varphi$, called "quintessence" [1] [2]

In the framework of a spatially flat Friedmann universe, $\Omega_x(z)$ is the normalized dark energy density as a function of scale factor which evolves as $\Omega_x(a) = \Omega_{x0} f(a) H_0^2 / H^2$ and

$$f(a) = exp\left[3 \int_1^a a(1 + \omega(a)) da\right], \qquad (2)$$

In order to characterize dark energy, there are many functions (Equations of State) describing and constraining dark energy's parameters. In this context we consider four popular parametrization.

A: CPL (Chevallier-Polarski-Linder) [3] [4]

$$\omega(a) = \omega_0 + \omega_a(1 - a) \qquad (3)$$

B: Jassal model [5]

$$\omega(a) = \omega_0 + \omega_a a(1-a) \tag{4}$$

C: Barboza & Alcaniz model [6]

$$\omega(a) = \omega_0 + \omega_a \frac{(1-a)}{2a^2 - a + 1} \tag{5}$$

D: Wetterich model [7]

$$\omega(a) = \frac{\omega_0}{1 - \omega_a \ln a} \tag{6}$$

We use the best fit of dark energy parameters in Ref. [8] and put them in all calculations and plots.

| Parametrization | $\Omega_{m0}$ | $w_0$ | $w_1$ |
|---|---|---|---|
| A | $0.28 \pm 0.01$ | $-1.09 \pm 0.08$ | $0.43^{+0.26}_{-0.31}$ |
| B | $0.28 \pm 0.01$ | $-1.03 \pm 0.10$ | $0.95^{+0.92}_{-0.84}$ |
| C | $0.28 \pm 0.01$ | $-1.08 \pm 0.07$ | $0.23^{+0.14}_{-0.17}$ |
| D | $0.28 \pm 0.01$ | $-1.09 \pm 0.08$ | $0.17^{+0.11}_{-0.12}$ |

**Table 1** Marginalized Result with 1σ Errors for Each Parametrization [8]

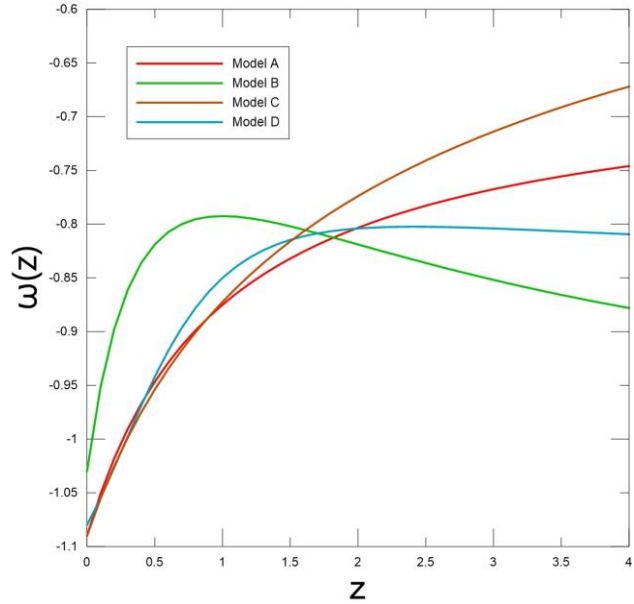

**Fig 1.** Evolution of EoS in terms of redshift. As it is evident, model A and C have almost identical behavior. Model B has rapid variation in low redshift

## Quintessence and Barotropic models

If we study the dynamic of quintessence on the flat *FLRW* universe with line element $ds^2 = -dt^2 + a^2(t)dx^2$, where *a(t)* is the scale factor in terms of time *t*, the pressure and energy density can be written as [9]

$$P_\phi = \frac{\dot{\phi}^2}{2} - V(\phi), \qquad (7)$$

$$\rho_\phi = \frac{\dot{\phi}^2}{2} + V(\phi), \qquad (8)$$

Where a dot represents a derivative with respect to *t*. Now we can rewrite equation (1) as

$$\omega \equiv \frac{P_\phi}{\rho_\phi} = \frac{\frac{\dot{\phi}^2}{2} - V(\phi)}{\frac{\dot{\phi}^2}{2} + V(\phi)}, \qquad (9)$$

The scalar field satisfies the continuity equation and equation of motion is [9]

$$\ddot{\phi} + 3H\dot{\phi} + V_{,\phi} = 0, \qquad (10)$$

Where $H = \dot{a}/a$ and $V_{,\phi} = dV/d\phi$. From the equations above, the scalar field potential in terms of energy of a perfect fluid can be obtained [10] [11]

$$V = \frac{1}{2}(1-\omega)\rho_\phi, \qquad (11)$$

And the density of perfect fluid is

$$\frac{a}{\rho}\frac{d\rho}{da} = -3(1+\omega), \qquad (12)$$

We apply this method to four typical parametrizations (similar to earlier works [12] [13]) which fit the data well [8] and discuss the general features of the resulting potentials.

A: CPL (Chevallier-Polarski-Linder)

$$\rho = \rho_{\phi 0} a^{-3(1+\omega_a+\omega_0)} e^{3\omega_a(a-1)}, \qquad (13)$$

$$V(a) = \frac{1}{2}\rho_{\phi 0}(1-\omega_0-\omega_a+\omega_a a)a^{-3(1+\omega_0+\omega_a)}e^{3\omega_a(a-1)}, \qquad (14)$$

B: Jassal model

$$\rho = \rho_{\phi 0} a^{-3(1+\omega_0)} e^{\frac{3}{2}\omega_a(a^2-2a-1)}, \qquad (15)$$

$$V(a) = \frac{1}{2}\rho_{\phi 0}(1 - \omega_0 - \omega_a a(1-a))a^{-3(1+\omega_0)} e^{\frac{3}{2}\omega_a(a^2-2a-1)} \qquad (16)$$

C: Barboza & Alcaniz model

$$\rho = \rho_{\phi 0} a^{-3(1+\omega_a+\omega_0)} \left(a^2 - a + \frac{1}{2}\right)^{\frac{3}{2}\omega_a} e^{1.03\omega_a}, \qquad (17)$$

$$V(a) = \frac{1}{2}\rho_{\phi 0}\left(1 - \omega_0 - \omega_a \frac{(1-a)}{2a^2 - a + 1}\right) a^{-3(1+\omega_a+\omega_0)} \left(a^2 - a + \frac{1}{2}\right)^{\frac{3}{2}\omega_a} e^{1.03\omega_a} \qquad (18)$$

D: Wetterich model

$$\rho = \rho_{\phi 0} a^{-3} e^{3\frac{\omega_0}{\omega_a}\left(\frac{1}{1-\omega_a \ln a} - 1\right)}, \qquad (19)$$

$$V(a) = \frac{1}{2}\rho_{\phi 0}\left(1 - \frac{\omega_0}{1 - \omega_a \ln a}\right) a^{-3} e^{3\frac{\omega_0}{\omega_a}\left(\frac{1}{1-\omega_a \ln a} - 1\right)}, \qquad (20)$$

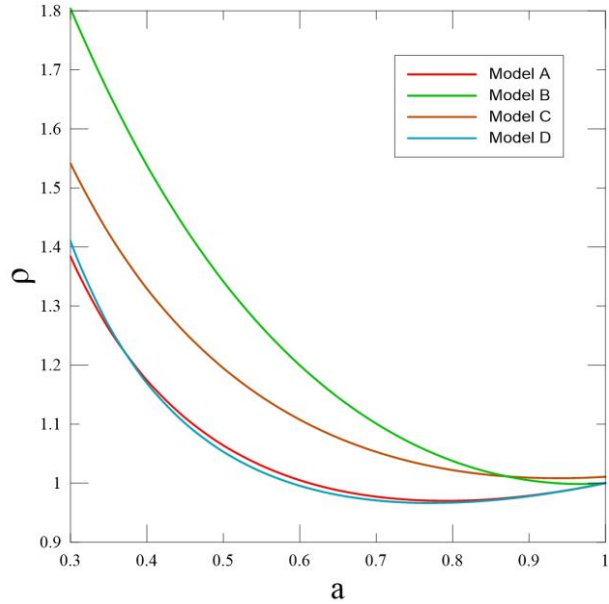

**Fig 2.** Evolution of the energy density of the quintessence. Model A and D have identical behavior and at scale factor 1 or redshift 0 (present time) model A, B and D have similar to each other but model C has a small difference.

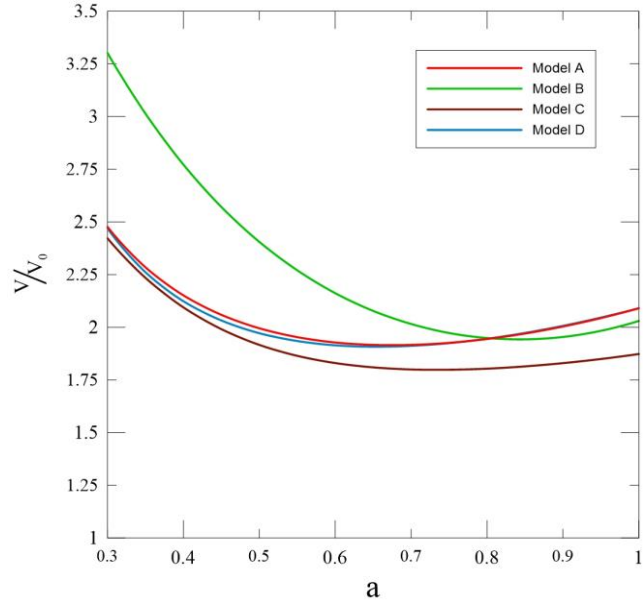

**Fig 3.** Evolution of normalized potential in terms of scale factor

Expression of $V$ in terms of $\phi$ leads us to obtain the following term using (7), (8), (9), (10) equations

$$\frac{d\phi}{dt} = [(1+\omega)\rho_\phi]^{\frac{1}{2}}, \tag{21}$$

One can rewrite the equation above from time dependence to an equation in terms of scale factor $a$

$$\frac{d\phi}{da} = \frac{[(1+\omega)\rho_\phi]^{\frac{1}{2}}}{aH}, \tag{22}$$

Where $H = [(\rho_\phi + \rho_M)/3]^{1/2}$ and $\rho_M = \rho_{M0} a^{-3}$.

Substituting four parametrization models for $\omega$ into equation (22) yields the following terms

$$\phi_A = \int \frac{\sqrt{3(1+\omega_0+\omega_1-\omega_1 a)}}{\sqrt{1+\frac{\rho_{M0}}{\rho_{\phi 0}} a^{3(\omega_0+\omega_1)} e^{3\omega_1(1-a)}}} \frac{da}{a}, \tag{23}$$

$$\phi_B = \int \frac{\sqrt{3(1+\omega_0+\omega_1 a(1-a))}}{\sqrt{1+\frac{\rho_{M0}}{\rho_{\phi 0}} a^{\omega_0} e^{-6\omega_1(a^2-2a+1)}}} \frac{da}{a}, \tag{24}$$

$$\phi_C = \int \frac{\sqrt{3(1+\omega_0+\omega_1\frac{1-a}{2a^2-2a+1})}}{\sqrt{1+\frac{\rho_{M0}}{\rho_{\phi0}}a^{(\omega_0+\omega_1)}(a^2-2a+\frac{1}{2})^{\frac{3}{2}\omega_1}e^{1.03\omega_1}}}\frac{da}{a}, \qquad (25)$$

$$\phi_D = \int \frac{\sqrt{3(1+\frac{\omega_0}{(1+\omega_1 lna)^2})}}{\sqrt{1+\frac{\rho_{M0}}{\rho_{\phi0}}a^{\omega_0}e^{-3\frac{\omega_0}{\omega_1}(\frac{1}{1-\omega_1 lna}-1)}}}\frac{da}{a}, \qquad (26)$$

In order to find out the behavior of $V(\phi)$ one can plot equations (11) and (22) together. Note that calculating equations (23)-(26) needs limited amounts of $\omega_0$ and $\omega_1$. Then instead of table 1, as a toy model, we consider -0.3 and -0.5 for $\omega_0$ and $\omega_1$ respectively in all models.

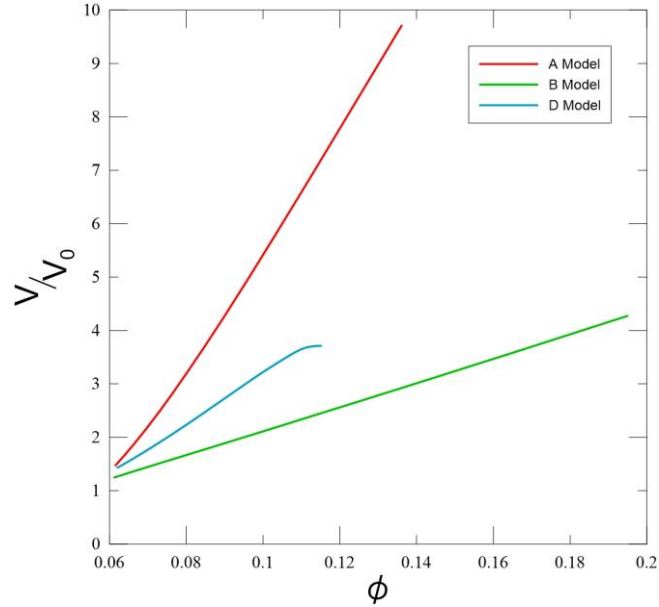

**Fig 4.** Describing normalized scalar field potential as a function of $\phi$ for model A, B and D. we just consider three models which are identical in terms of their total value. Model C is neglected for this figure.

For barotropic model we have the pressure as a function of the density [14]

$$P = f(\rho) \qquad (27)$$

If we obey the constraints of prior work for barotropic models [15], it is better to derive pressure as a function of the density and use the limitation for $dp/d\rho$

$$0 < \frac{dp}{d\rho} < 1 \tag{28}$$

Substituting energy density of four mentioned parametrizations into $P = \omega\rho$ results in the following equations for pressure

$$P_A = \rho_{\phi 0}[\omega_0 + \omega_a(1-a)]a^{-3(1+\omega_a+\omega_0)}e^{3\omega_a(a-1)}, \tag{29}$$

$$P_B = \rho_{\phi 0}[\omega_0 + \omega_a a(1-a)]a^{-3(1+\omega_0)}e^{\frac{3}{2}\omega_a(a^2-2a-1)}, \tag{30}$$

$$P_C = \rho_{\phi 0}\left[\omega_0 + \omega_a\frac{(1-a)}{2a^2-a+1}\right]a^{-3(1+\omega_a+\omega_0)}\left(a^2-a+\frac{1}{2}\right)^{\frac{3}{2}\omega_a}e^{1.03\omega_a}, \tag{31}$$

$$P_D = \rho_{\phi 0}\left[\frac{\omega_0}{1-\omega_a\ln a}\right]a^{-3}e^{3\frac{\omega_0}{\omega_a}\left(\frac{1}{1-\omega_a\ln a}-1\right)}. \tag{32}$$

Now, plotting both equations (12) and $P = \omega\rho$ together provide an expression for $f(\rho)$.

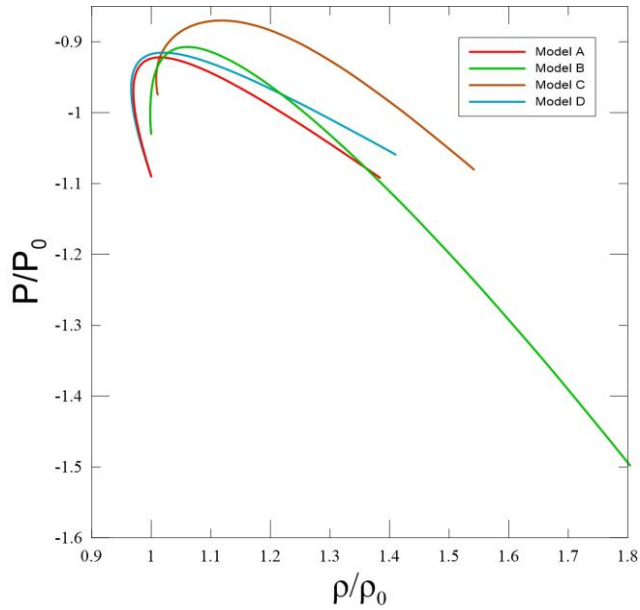

**Fig 5.** For barotropic models described by the four models, the pressure p, given as $p/p_0$, is plotted as a function of the density $\rho$, given as $\rho/\rho_0$, where $\rho_0$ is the present-day dark energy density. All models approximately reach one value at 1.05 for density

**Conclusion**

Despite that in fig.1 all models except model B have almost the same behaviors, model A and C have more similarity in there trends. On the other hand, figures 2 and 3 and also fig. 5 demonstrate that model A and D are more similar to each other and model C has a different variation. In better word model D can be a good substitution for most used CPL parametrization. Although, trends of these models in $V(\phi)$ are different, but figure 5 in barotropic model shows a good fit for model A and D. This gives a good result, Wetterich Model as a nonlinear model similar to CPL as a linear approximation for equation of state parameter as a function of expansion is very useful.